\documentclass[9pt]{article}
\usepackage{authblk}
\usepackage{textcomp}
\usepackage{abstract}
\usepackage{graphicx}
\usepackage[utf8]{inputenc}
\usepackage[margin=1in]{geometry}
\usepackage[shortcuts]{extdash}

\title{Self-referenced frequency combs using high-efficiency silicon-nitride waveguides}

\author[1,*]{David R. Carlson}
\author[1]{Daniel D. Hickstein}
\author[1,2]{Alex Lind}
\author[3]{Stefan Droste}
\author[4]{Daron Westly}
\author[3]{Nima Nader}
\author[3]{Ian Coddington}
\author[3]{Nathan R. Newbury}
\author[4]{Kartik Srinivasan}
\author[1,2]{Scott A. Diddams}
\author[1,2]{Scott B. Papp}

\affil[1]{Time and Frequency Division, National Institute of Standards
  and Technology, 325 Broadway, Boulder, CO, 80305}

\affil[2]{Dept. of Physics, University of Colorado, 2000 Colorado Ave, Boulder,
  CO, 80309}

\affil[3]{Applied Physics Division, National Institute of Standards
  and Technology, 325 Broadway, Boulder, CO, 80305}

\affil[4]{Center for Nanoscale Science and Technology, National
  Institute of Standards and Technology, 100 Bureau Drive, Gaithersburg, Maryland, 20899}

\affil[*]{Corresponding author: david.carlson@nist.gov}

\date{}

\begin{document}

\maketitle

\begin{abstract}
We utilize silicon-nitride waveguides to self-reference a telecom-wavelength
fiber frequency comb through supercontinuum generation, using 11.3~mW of optical power incident on the chip. This is approximately ten times lower than conventional approaches using nonlinear fibers and is enabled by low-loss ($<$2~dB) input coupling and the high nonlinearity of silicon nitride, which can provide two octaves of spectral broadening with incident energies of only 110 pJ. Following supercontinuum generation, self-referencing is accomplished by mixing 780\=/nm dispersive-wave light with the frequency-doubled output of the fiber laser. In addition, at higher optical powers, we demonstrate  $f$\=/to\=/$3f$ self-referencing directly from the waveguide output by the interference of simultaneous supercontinuum and third harmonic generation, without the use of an external doubling crystal or interferometer. These hybrid comb systems combine the performance of fiber-laser frequency combs with the high nonlinearity and compactness of photonic waveguides, and should lead to low-cost, fully stabilized frequency combs for portable and space-borne applications.

\end{abstract}

\renewcommand{\floatpagefraction}{.8}

%\section{Introduction}

Chip-integrated photonic waveguides, due to their high spatial light confinement and strong nonlinear response, are ideally suited for performing nonlinear optics with femtosecond laser pulses.  In particular, waveguide-based supercontinuum generation (SCG) using mode-locked laser frequency combs can produce broadband comb spectra spanning up to several hundred terahertz in a variety of different material platforms including silica~\cite{oh_supercontinuum_2014, oh_coherent_2017}, silicon-on-insulator~\cite{hsieh_supercontinuum_2007, kuyken_octave-spanning_2015}, AlGaAs~\cite{dolgaleva_broadband_2010}, chalcogenide glasses~\cite{eggleton_chalcogenide_2011}, aluminum nitride (AlN)~\cite{hickstein_ultrabroadband_2017}, and silicon nitride (Si$_3$N$_4$, henceforth SiN)~\cite{halir_ultrabroadband_2012, chavez_boggio_dispersion_2014, zhao_visible--near-infrared_2015, epping_-chip_2015}. In addition to possessing high nonlinearity, the SiN platform is especially attractive for many applications because of its compatibility with standard silicon fabrication techniques as well as having a broad transparency window extending from the visible to the mid-infrared. For example, these characteristics have recently allowed SiN waveguides to produce tailored two-octave output spectra suitable for precision frequency metrology~\cite{carlson_photonic-chip_2017}.

In addition to generating broad bandwidth, SiN offers an attractive way to generate the comb offset $f_0$.  Offset-frequency stabilization is most commonly accomplished using a self-referencing technique called $f$-to-$2f$ interferometry~\cite{jones_carrier-envelope_2000} that, for any fiber laser system, requires spectral broadening in a nonlinear fiber. Such nonlinear fibers require high peak powers in order to generate sufficient bandwidth, and consequently optical amplifiers are often needed to increase the pulse energy prior to broadening. On the other hand, the high nonlinearity and tight confinement of SiN allows much lower energies to be used for $f$-to-$2f$ broadening while additionally having a more compact form-factor~\cite{mayer_frequency_2015, klenner_gigahertz_2016}. Furthermore, SiN can support spectral broadening to twice the pump frequency to allow decoupling of the nonlinear broadening and frequency-doubling processes for reduced sensitivity to external perturbations~\cite{carlson_photonic-chip_2017, oh_coherent_2017}.

Offset-frequency stabilization of 1550~nm telecom-wavelength combs using nonlinear waveguides is of particular interest because these comb systems are well-developed, have low-cost components readily available, and are successfully used in many diverse fields. However, field-portable, space-borne, and integrated applications for these combs have very tight power and cost budgets. In such cases, a simple and low-power self-referencing solution is required~\cite{hartl_integrated_2005, ishizawa_demonstration_2008, jiang_tapered_2014, lezius_space-borne_2016}. Waveguide-broadened combs operating in the low-pulse-energy regime are also important as they offer a path toward the stabilization of high repetition rate electro-optic combs and fully chip-integrated microresonator combs. 

In this Letter, we present two approaches to self-referencing frequency combs with stoichiometric SiN waveguides that demonstrate the flexibility of this chip-integrated approach.  First, we demonstrate a waveguide design with improved input-coupling that enables $f_0$ stabilization with less than 150~pJ of pulse energy (15~mW total power, including for second-harmonic generation), a level attainable directly from fiber laser oscillators. Second, we show that SiN waveguides pumped with higher pulse energies can achieve simultaneous SCG and third-harmonic generation (THG) and enable $f$-to-$3f$ self-referencing directly from the waveguide. The resulting $2f_0$ beat frequency is used for comb-offset stabilization without the use of an external interferometer or a nonlinear doubling medium -- greatly simplifying the standard $f$-to-$2f$ technique and reducing the overall cost of the system.

\begin{figure}
\centering
\includegraphics[width=0.7\linewidth]{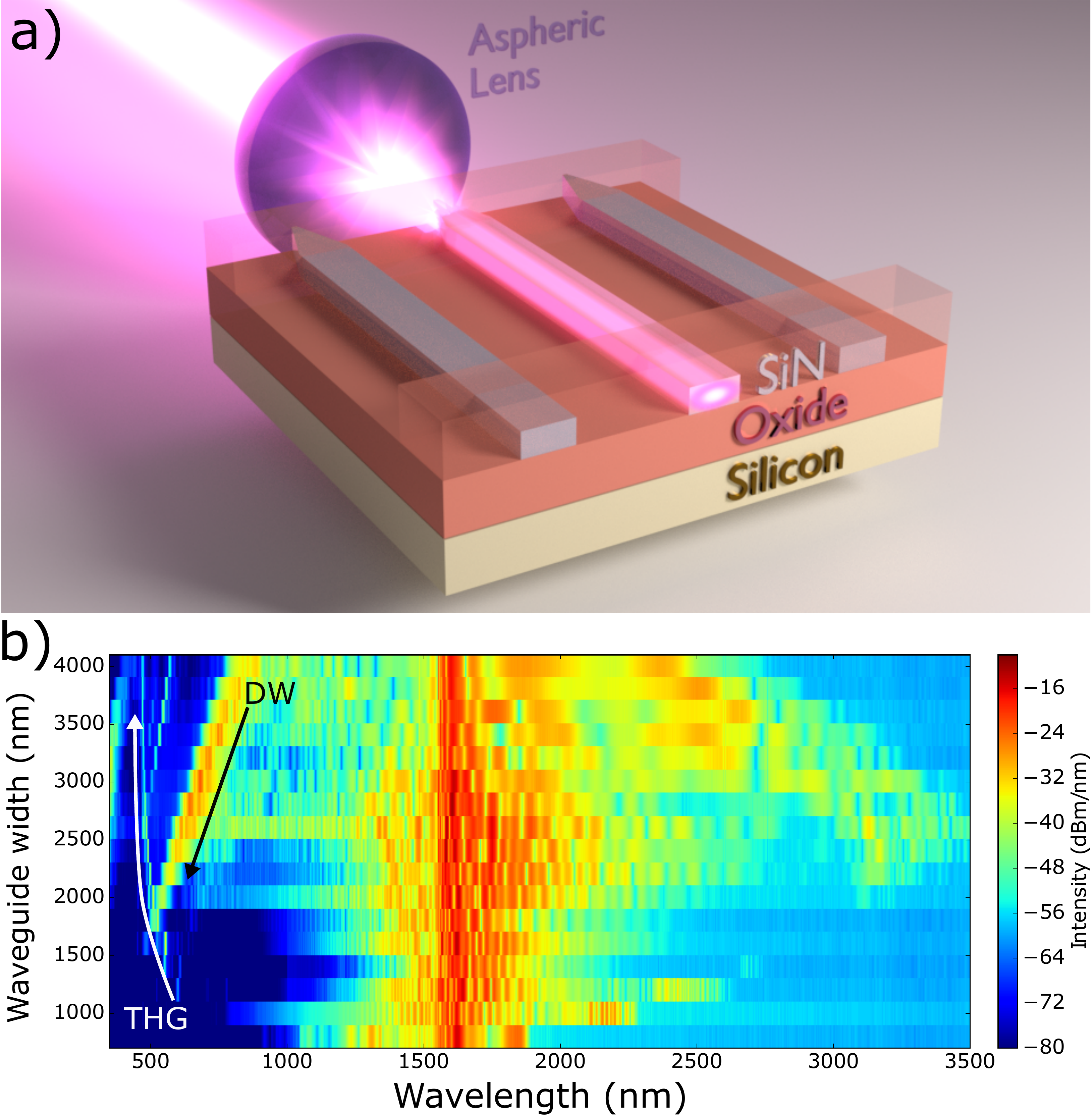}
\caption{a) Air-clad silicon nitride (SiN) waveguides with 700~nm thickness, 1~cm length, and varying widths. An inverse taper geometry at the input facet yields coupling loss less than 2~dB, while an oxide over-cladding at both input and output facets improves mode matching with free space optics. b) Supercontinuum spectra from SiN waveguides of different waveguide widths pumped with a 100~MHz repetition rate, 1550~nm fiber-laser frequency comb (DW: dispersive wave, THG: third-harmonic generation). For these spectra, the pulse energy was 1~nJ, the full-width at half-maximum pulse duration was 80~fs, and the total average power was 100~mW.}
\label{fig:highpowerspectrum}
\end{figure}

%\section*{Supercontinuum generation}
\begin{figure}
\centering
\includegraphics[width=0.7\linewidth]{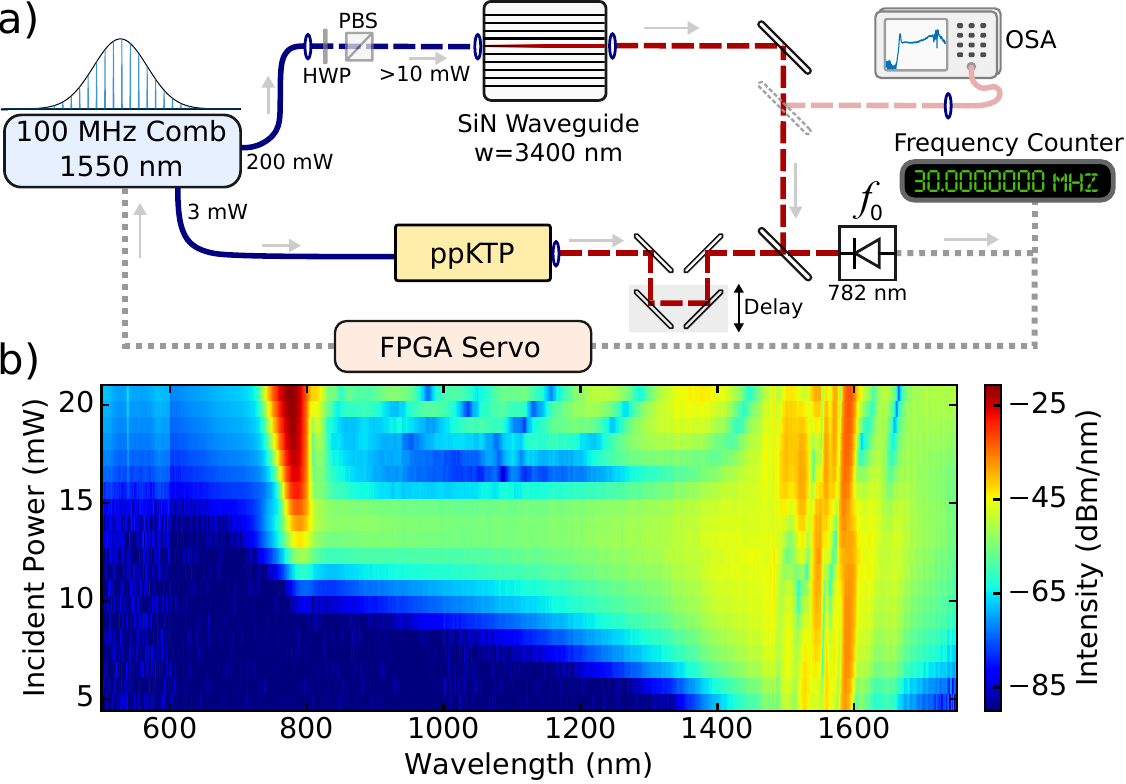}
\caption{a) Schematic of low-power self-referencing scheme (fiber path: solid lines, free-space: dashed lines, electrical path: dotted gray lines, HWP: half-wave plate, PBS: polarizing beam-splitter, OSA: optical spectrum analyzer). A turn-key fiber-laser frequency comb is coupled into a silicon nitride waveguide.  The waveguide output is collected with a 0.85~NA microscope objective and overlapped with doubled pump light from a periodically-poled potassium titanyl phosphate (ppKTP) waveguide to obtain an $f_0$ signal after photodetection. The offset frequency is locked using a digital FPGA-based servo loop by feeding back to the fiber laser pump current. A flipper mirror can be used to divert the waveguide output light to a multimode fiber for recording the optical spectrum as a function of average incident laser power as shown in b). A dispersive wave at 775~nm is observed with at least 10~mW of incident power.}
\label{fig:lowpowerspectrum}
\end{figure}

The 1~cm long chip-integrated waveguides used in this work have an air-clad ridge geometry
(see Fig.~\ref{fig:highpowerspectrum}a) and are made of 700~nm thickness low pressure
chemical vapor deposition (LPCVD) stoichiometric SiN.  The waveguide pattern is written to the chip using electron-beam lithography and includes inverse tapers at the input facet to improve optical coupling~\cite{cardenas_high_2014}. Though several techniques have been demonstrated to offer very high efficiency fiber-to-waveguide coupling~\cite{tiecke_efficient_2015, groblacher_highly_2013, chen_low-loss_2010},  free-space coupling is used here for convenience.  An input coupling loss of less than 2~dB is achieved using an aspheric lens with a design wavelength of 1550~nm and numerical aperture (NA) of 0.6. In addition to the inverse taper region that adiabatically expands the mode, there is an oxide over-cladding near the edges of the chip to improve the mode symmetry and overlap with the incident beam. However, the output of the waveguide is not tapered to achieve consistent output coupling across the spectrum and to avoid additional absorption of the long-wavelength spectral components in the fully oxide-clad region.

The frequency comb source is an amplified 1550-nm mode-locked all-polarization-maintaining fiber laser producing 80~fs FWHM sech$^2$ pulses at an average power of 200~mW ($f_r = 100$~MHz)~\cite{sinclair_invited_2015}. For these tests, its output is attenuated using a half-wave-plate and polarizing beam-splitter before coupling to the waveguide. We operate the amplifier at higher power than is necessary for SCG in the waveguide in order to maintain optimal pulse compression. However, such short pulses could be obtained directly from an appropriately configured laser oscillator without an external amplifier \cite{zhang_sub-100_2013, li_direct_2015}.

The SC spectrum for an incident pulse energy of 1~nJ is shown in Fig.~\ref{fig:highpowerspectrum}b as a function of waveguide width. Light covering more than two octaves of bandwidth is generated from approximately 500~nm to beyond 3~\textmu{}m and is similar to the SiN spectra recently reported in~\cite{porcel_two-octave_2017}. For self-referencing, the two relevant features of the supercontinuum are the short-wavelength dispersive wave and third-harmonic generated light, as discussed in the following sections. 

\section*{Low-power comb offset stabilization}
For $f$-to-$2f$ interferometry, we exploit the dispersive wave in the 782~nm spectral region that is generated by the 3400~nm wide waveguide, as shown in Fig.~\ref{fig:highpowerspectrum}b. This light can then be heterodyned against doubled light from the 1550-nm fiber-laser pulses to generate the offset frequency. In contrast to the traditional approach in highly nonlinear fiber of achieving an octave-spanning spectrum and then mixing doubled 2~\textmu{}m light with fundamental 1~\textmu{}m light, in this method we have effectively decoupled the SCG from the frequency doubling. This allows for more relaxed requirements on the SCG and reduces the sensitivity to spectral power fluctuations.

\begin{figure}
\centering
\includegraphics[width=0.7\linewidth]{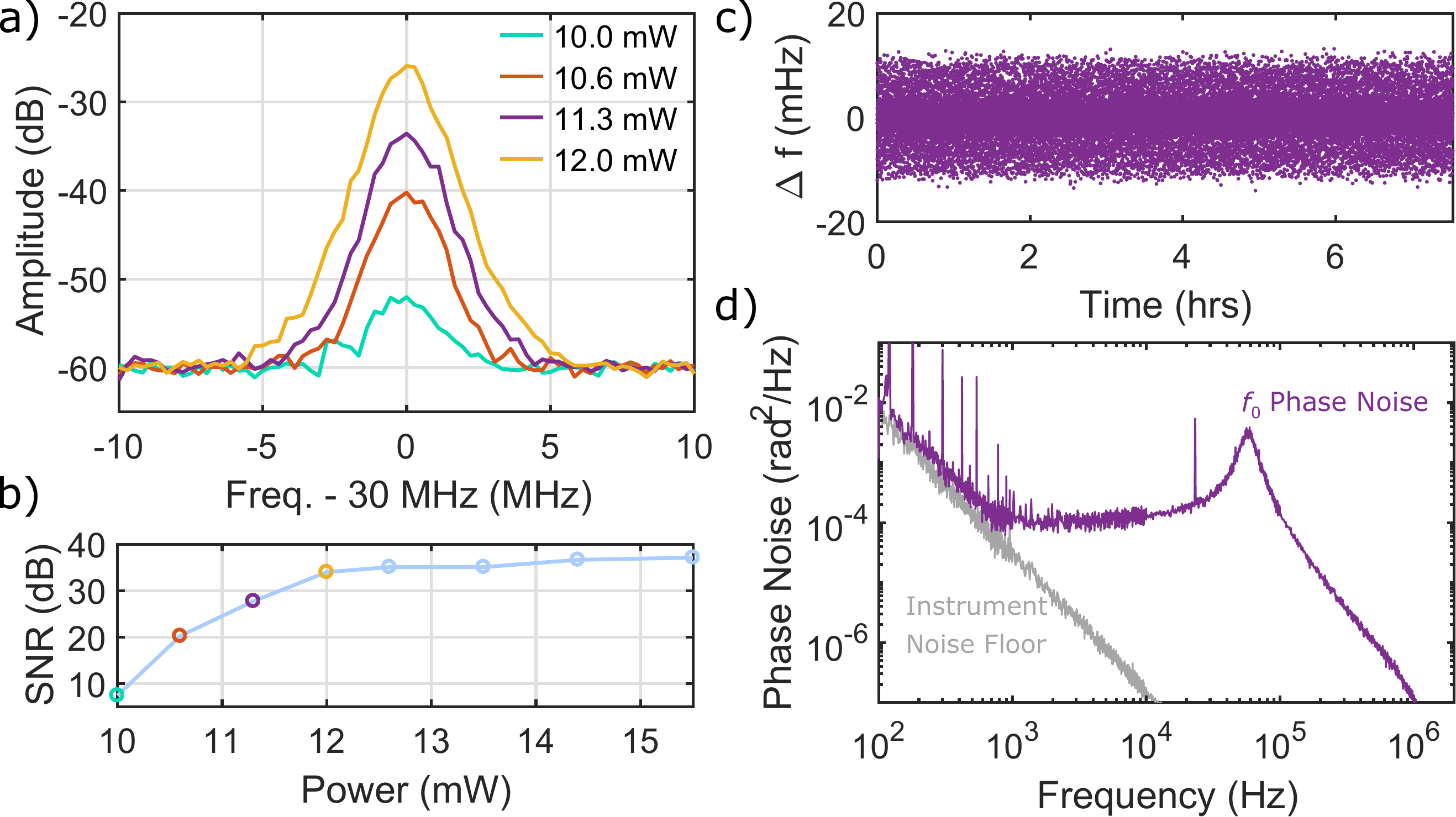}
\caption{a) Locked offset-frequency and b) corresponding SNR as a function of incident laser power at 1~MHz RBW. c) In-loop frequency deviation $\Delta f = f_0 - 30$ MHz as recorded by an external frequency counter (1~s gate time) while $f_0$ is locked using 11.3~mW incident optical power. The comb is locked without cycle slips for the full 7.5~hour acquisition. d) Phase noise spectrum for the locked $f_0$ beat note.}
\label{fig:rfbeats}
\end{figure}

Detection and subsequent stabilization of the comb offset frequency at 782~nm is accomplished using the schematic setup shown in Fig.~\ref{fig:lowpowerspectrum}a. The  SC spectra obtained from the 3400~nm width waveguide at low incident average powers is shown in Fig.~\ref{fig:lowpowerspectrum}b.  The sharp onset of dispersive wave generation at 10~mW incident power indicates the threshold for soliton fission~\cite{dudley_supercontinuum_2006}. An f-2f interferometer combines this dispersive wave with doubled light generated by diverting a small amount (3~mW) of the comb power to a periodically-poled fiber-coupled potassium titanyl phosphate (ppKTP) waveguide.  The detected $f_0$ beat note is shown in Fig.~\ref{fig:rfbeats}a for different incident powers to the waveguide. A digital field-programmable gate array (FPGA) servo loop filter digitizes and electronically filters the offset frequency signal before applying a correction signal to the fiber laser's pump with approximately 60~kHz bandwidth~\cite{sinclair_invited_2015}.  

%Unlike analog control electronics, this digital control digital  is able to track large phase excursions and has more %flexibility in tuning both filters and loop parameters. \textit{could delete this last sentence as not the main poitn %of the paper}

For an out-of-loop verification that the comb-offset lock is performing as expected, we count the locked offset frequency with a standard high-resolution commercial frequency counter ($\Lambda$-type). This counter requires greater than 25~dB signal-to-noise ratio (SNR) in a 1~MHz resolution bandwidth (RBW), which is achieved here for incident laser power onto the waveguide as low as 11.3~mW.   Fig.~\ref{fig:rfbeats}c shows the counter record of the stabilized in-loop offset frequency at 30~MHz for 7.5~hours of continuous operation. No cycle slips are observed within this period, indicating a robust phase lock is achieved. At an averaging time of $\tau = 1$~s, the in-loop frequency instability is $1\times10^{-17}$ and averages down at a rate of $\tau^{-1/2}$. The integrated RMS phase fluctuation of the locked offset frequency is approximately 6~rad (from 1 Hz to 1 MHz, see phase noise in Fig.~\ref{fig:rfbeats}d), and is set by the noise properties of this soliton laser design and feedback bandwidth, and not the $f$-to-$2f$ detection~\cite{sinclair_invited_2015}. To the best of our knowledge, this is the lowest total average power used to self-reference a frequency comb through nonlinear broadening.

\section*{\lowercase{\emph{f}-to-3\emph{f}} comb offset stabilization }
\begin{figure}
\centering
\includegraphics[width=0.7\linewidth]{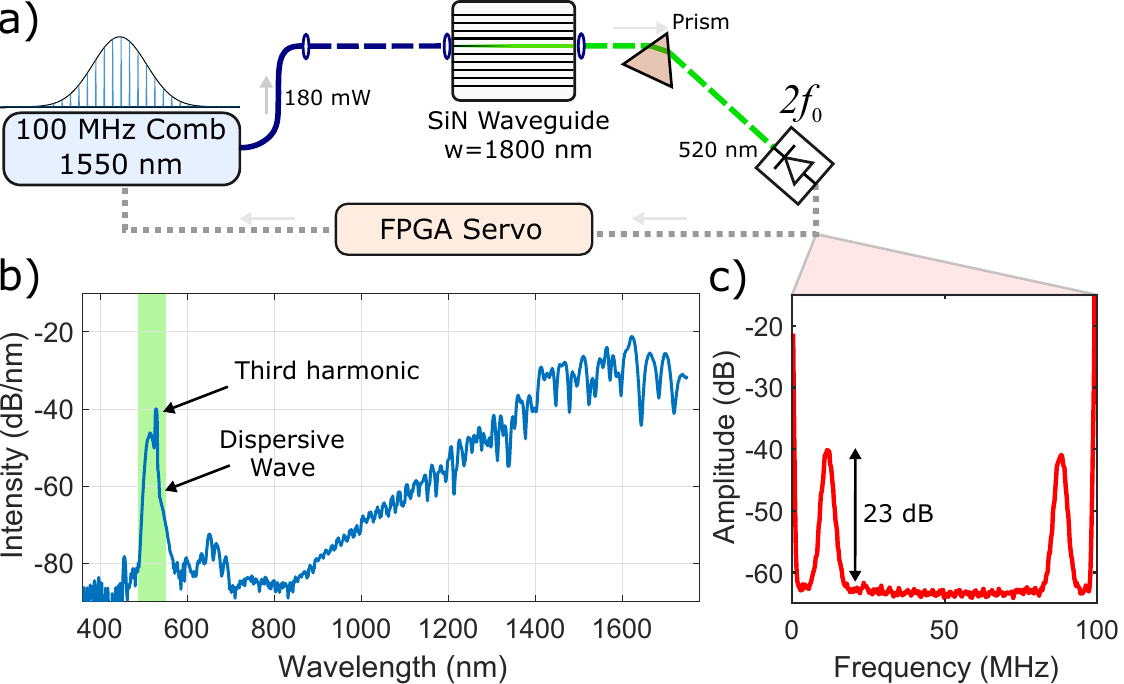}
\caption{a) Schematic for $f$-to-$3f$ self-referencing (fiber path: solid lines,
  free-space: dashed lines, electrical path: dotted gray lines). The waveguide output is collimated and spectrally resolved using a prism before
  illuminating a photodetector.  b) Supercontinuum spectrum of the 1800-nm wide
  waveguide from Fig.~\ref{fig:highpowerspectrum} showing overlapped
  third-harmonic and dispersive wave contributions. The highlighted region is
  photodetected to obtain a beat frequency for $f$-to-$3f$ self-referencing. c)
  RF signal at $2f_0$, obtained directly from the SiN waveguide output. The
  signal, with a signal-to-noise ratio of 23~dB at 1~MHz RBW, is suitable for
  locking using a feedback servo.}
\label{fig:f3f}
\end{figure}

For applications where simplicity takes priority over optical power consumption, we demonstrate an alternative self-referencing scheme using SiN waveguides that eliminates the external doubling crystal as well as the interferometer. This technique is similar to the recently reported $f$-to-$2f$ stabilization directly from the output of aluminum nitride (AlN) waveguides, which exploited the simultaneous presence of both second-order $\chi^{(2)}$ and third-order $\chi^{(3)}$ nonlinearities~\cite{hickstein_ultrabroadband_2017}. However, since SiN is centrosymmetric and thus only supports $\chi^{(3)}$, only third harmonic light can be generated and we instead require supercontinuum broadening to $3f$ .  In this case, the resulting beat note from the interference between comb modes $3\nu_n$ and  $\nu_{3n}$ due to simultaneous THG and SCG, respectively, occurs at frequency
\begin{eqnarray}
\label{eq:f3f}
	3\nu_n - \nu_{3n} &=& 3(nf_r + f_0) - (3nf_r + f_0)\quad  =\quad 2 f_0.
\end{eqnarray}

The spectra displayed in Fig.~\ref{fig:highpowerspectrum}b show that for a waveguide width of 1800~nm and 1~nJ input pulse energy, the THG peak and supercontinuum dispersive wave intersect near 520~nm.  In this spectral region it is possible to both observe and lock the $2f_0$ beat frequency from Eq.~\ref{eq:f3f}.  A schematic of the detection scheme and spectrum from the waveguide are shown in Fig.~\ref{fig:f3f}. The collimated waveguide output is spectrally filtered using a prism and then spatially filtered by a small-area avalanche-photodetector to find the highest degree of mode overlap. 

\begin{figure}
\centering
\includegraphics[width=0.7\linewidth]{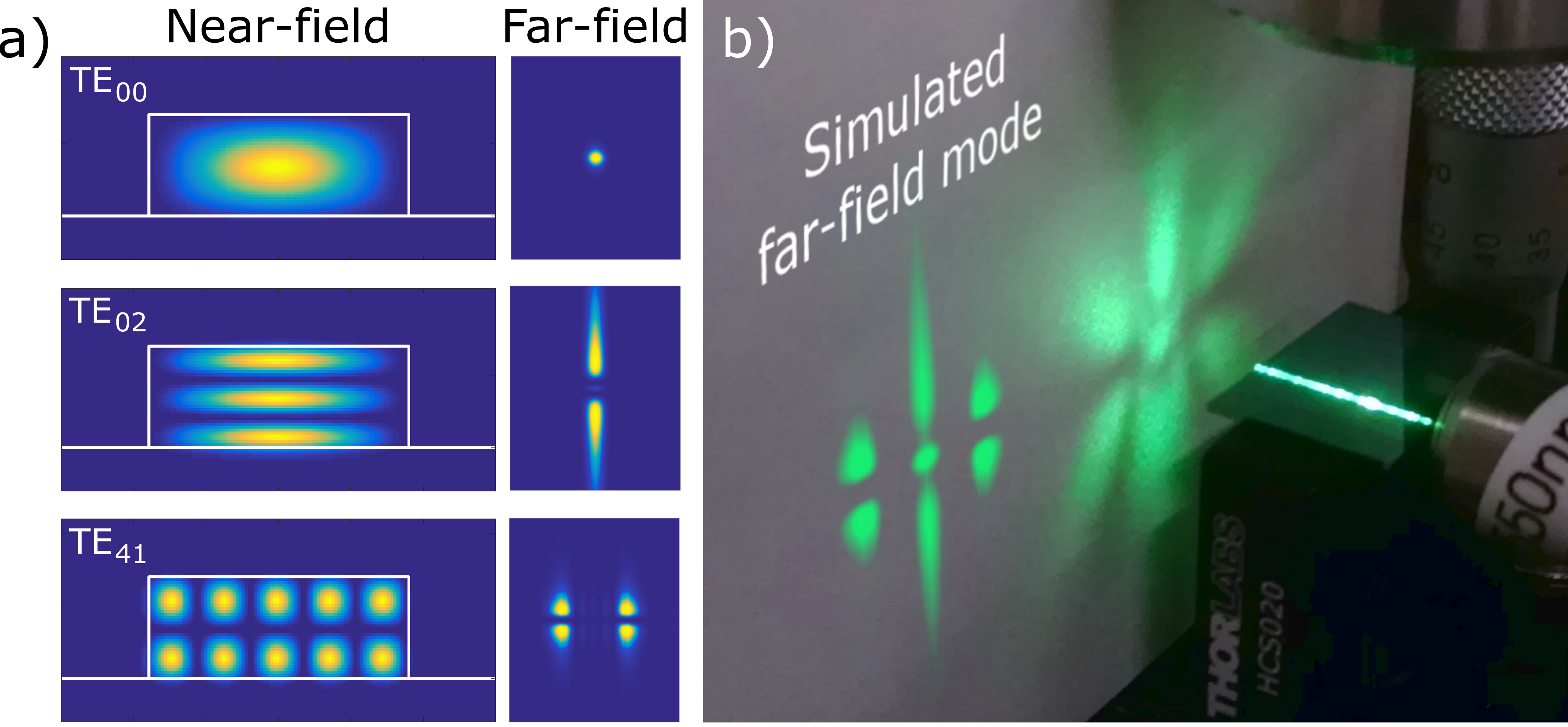}
\caption{a) Simulated near-field and far-field intensity profiles at a wavelength of 520~nm for the modes of the 1800-nm width waveguide contributing to $f$-to-$3f$ self-referencing.  The calculations were performed using the \texttt{wgmodes} mode solver~\cite{fallahkhair_vector_2008}. b) Experimental (right) and simulated (left) output mode obtained from a rectilinear projection of the combined modes in a). Poor mode overlap limits the achievable beat note SNR.}
\label{fig:f3fmodes}
\end{figure}

Though it is not readily visible in the optical spectrum, there are two different phase-matched THG modes in the photodetected region that can contribute to the $f$-to-$3f$ signal. Unfortunately, these higher order modes (TE$_{02}$ and TE$_{41}$, see Fig.~\ref{fig:f3fmodes}) only exhibit very small spatial overlap (i.e. heterodyne mixing efficiency) with the fundamental in both the near- and far-field, limiting the achievable SNR. Temporal overlap between the $f$ and $3f$ spectral components is also a concern as the narrow-bandwidth spectrum of the third harmonic leads to a longer pulse and the waveguide modal dispersion can lead to temporal walk-off.  Nevertheless, the digital FPGA servo used in the system is able to lock to the optimized 23~dB SNR $2f_0$ signal with an integrated phase noise of approximately 13~mrad.  As expected from Eq.~\ref{eq:f3f}, this value is roughly twice the 6~mrad RMS phase noise obtained with the low-power $f_0$ stabilization scheme in the previous section, suggesting a high-quality lock. However, due to insufficient SNR, it is not possible to verify this stabilization with an external frequency counter.  In the future, an on-chip mode converter~\cite{guan_ultra-compact_2015} could be used to isolate a single third-harmonic mode and achieve optimal spatial and temporal overlap for increasing the SNR of the $2f_0$ beat.

% include a few summary sentences. 

The two self-referencing techniques described in this work highlight the flexibility of the SiN material platform for supporting frequency comb systems. The low power consumption and experimental simplicity that are achievable with photonic integration will enable the widespread use of combs outside of the laboratory in applications ranging from fieldable dual-comb spectroscopy systems, to space-based optical clocks, as well as to remote sensing.

%\section*{Funding Information}

This research is supported by the Air Force Office of Scientific Research
(AFOSR) under award number FA9550-16-1-0016, the Defense Advanced Research
Projects Agency (DARPA) ACES program, the National Aeronautics and Space
Administration (NASA), the National Institute of Standards and Technology
(NIST), and the National Research Council. This work is a contribution of the U.S. government and is not subject to
copyright.

%\section*{Acknowledgments}
% * Certain commercial equipment, instruments, or materials are identified in this
% paper in order to specify the experimental procedure adequately. Such
% identification is not intended to imply recommendation or endorsement by the
% National Institute of Standards and Technology, nor is it intended to imply that
% the materials or equipment identified are necessarily the best available for the
% purpose.

% Bibliography
\bibliography{Zotero}
\bibliographystyle{osajnl}

% Full bibliography added automatically for Optics Letters submissions
% Note that this extra page will not count against page length
% \ifthenelse{\equal{\journalref}{ol}}{%
% \clearpage
% \bibliographyfullrefs{Zotero}
% }{}

\end{document}